\documentclass{aa}  
\usepackage{graphicx}
\usepackage{txfonts}
\usepackage{lscape}

\usepackage{natbib}
\bibpunct{(}{)}{;}{a}{}{,} 

\begin{document}
   \title{A cool starspot or a second transiting planet in the TrES-1 system?\thanks{Based on observations made with the NASA/ESA Hubble Space Telescope, obtained at the Space Telescope Science Institute, which is operated by the Association of Universities for Research in Astronomy, Inc., under NASA contract NAS 5-26555.}}

   \subtitle{}

   \author{M. Rabus\fnmsep
          \inst{1}
	  \and
	  R. Alonso\inst{2}
	  \and
	  J. A. Belmonte\inst{1}
	  \and
	  H. J. Deeg\inst{1}
	  \and
	  R. L.Gilliland\inst{3}
	  \and
	  J. M. Almenara Villa\inst{1}
	  \and
	  T. M. Brown\inst{4}
	  \and
	  D. Charbonneau\inst{5}
	  \and
	  G. Mandushev\inst{6}
          }

   \offprints{mrabus@iac.es}

   \institute{Instituto de Astrof\'{i}sica de Canarias,
              V\'ia Lactea s/n, E-38205 La Laguna, Spain\\
              \email{mrabus@iac.es}
         \and
             Laboratoire d'Astrophysique de Marseille (UMR 6110),
	     Technop\`ole de Marseille-Eoile, F-13388 Marseille, France\\
         \and
	     Space Telescope Science Institute, 
             3700 San Martin Drive, Baltimore, MD 21218, USA\\
	 \and
             Las Cumbres Observatory,
	     6740 Cortona Dr. Suite 102, 93117 Santa Barbara, USA\\
	 \and
             Harvard-Smithsonian Center for Astrophysics,
	     60 Garden Street, Cambridge, MA 02138, USA\\
	 \and
             Lowell Observatory,
	     1400 West Mars Hill Road, Flagstaff, AZ 86001, USA\\
             }

   \date{Received ; accepted }

 
  \abstract
   {}
   {We investigate the origin of a flux increase found during a transit of TrES-1, observed with the HST. This feature in the HST light curve cannot be attributed to noise and is supposedly a dark area on the stellar surface of the host star eclipsed by TrES-1 during its transit. We investigate the likeliness of two possible hypothesis for its origin: A starspot or a second transiting planet.}
   {We made use of several transit observations of TrES-1 from space with the HST and from ground with the IAC-80 telescope. On the basis of these observations we did a statistical study of flux variations in each of the observed events, to investigate if similar flux increases are present in other parts of the data set.}
   {The HST observation presents a single clear flux rise during a transit  whereas the ground observations led to the detection of two such events but with low significance.  In the case of having observed a starspot in the HST data, assuming a central impact between the spot and TrES-1, we would obtain a lower limit for the spot radius of 42000 km. For this radius the spot temperature would be 4690 K, 560 K lower then the stellar surface of 5250 K. For a putative second transiting planet we can set a lower limit for its radius at 0.37 R$_J$ and for periods of less than 10.5 days, we can set an upper limit at 0.72 R$_J$.}
   {Assuming a conventional interpretation, then this HST observation constitutes the detection of a starspot. Alternatively, this flux rise might also be caused by an additional transiting planet. The true nature of the origin can be revealed if a wavelength dependency of the flux rise can be shown or discarded with a higher certainty. Additionally, the presence of a second planet can also be detected by radial velocity measurements. }

   \keywords{starspots -- 
             planetary systems
               }

   \maketitle
%

\section{Introduction}

TrES-1 was the first planet transiting a bright star that was discovered by a wide-field survey \citep{2004ApJ...613L.153A}, and it has become a frequently observed and well characterized exoplanet. It orbits with a period of 3.03 days a bright K0V star with an apparent magnitude of V=11.79, approximately 157 pc away from us. A detailed light curve analysis done by \citet{2007ApJ...657.1098W} allowed to lower the errors of the star/planet parameters. \citet{2005ApJ...626..523C} obtained the first direct detection of light emitted from an exoplanet using the Spitzer Space Telescope by an observation of the TrES-1 secondary eclipse. \citet{2004ApJ...616L.167S,2006AJ....131.2274S} analyzed the chemical composition of the host star and compared it with other known planet-hosting stars. They found no chemical peculiarity compared to the other stars. \\

With the aim to acquire a very precise transit light curve of TrES-1, our group observed in the year 2005 several transit events with the Hubble Space Telescope (GO-10441). The most relevant finding has been a flux increase during one transit, which is clearly a feature coming from the TrES-1 system itself, for a preliminary description see \citet{2007prpl.conf..701C}. Coverage of numerous further transits events was obtained between 2004 and 2007 with the IAC 80-cm telescope (IAC-80). The motivation for this coverage was for one a long-term study of transit-timing variations and later, a follow-up of the HST observations. The analysis of the flux-rises detected in these observations constitutes the main part of this paper, whereas the results from the timing study will be reported elsewhere. \\

Both on TrES-1 \citep{2007prpl.conf..701C} and on other systems (HD209458, by \citeauthor{2003ApJ...585L.147S} 2003 based on data from \citeauthor{2001NewA....6...51D} 2001; HD189733 by \citeauthor{2007A&A...476.1347P} 2007), there have been reports of potential stellar spots detected from flux-rises during planetary transits. In starspots, intense magnetic fields can suppress the convection of heat to the surface, and hence, that area will radiate less light, causing a dark area on the stellar surface. Starspots could cover up to 55 \% of the stellar surface \citep{1998ApJ...501L..73O} and are generally 500 to 2000 Kelvin cooler then the surrounding area. The possibility to detect these during transits was first discussed by  \citet{2003ApJ...585L.147S}. Their simulations showed that a starspot fully or partly occulted by the transiting planet leaves a short brightness increase in the light curve of a transit. The flux variation feature from a starspot depends on two parameters: Its duration depends on the transit path-length across the spot and across the stellar surface, on the rotational velocity of the parent star, and on the projected orbital speed of the transiting planet. The ratio of transit path-lengths may be interpreted in first order, assuming a central impact,  as an area ratio between spot and the transiting planet \citep{2003ApJ...585L.147S}. The feature's amplitude for another, depends on the temperature difference between the normal stellar surface and the spot as well as on the fraction of the planet's shadow that grazes over the spot, unless the projected planet comes to lie fully within the starspot, in which case only the temperatures are relevant.\\

For example, \citet{2007A&A...476.1347P} found clear evidence of starspots on the star HD189733 by observing a transit of its planet with the HST at a high signal-to-noise ratio. They found an increase in flux of 0.001 mags and attributed this to the planet occulting a starspot. On TrES-1, there has been a previous ground-based search for spots during transits by \citet{2007ApJ...657.1098W} with negative results. They gave an upper 2 $\sigma$ limit for spots of $f < 4 x 10^{-4}$, where $f$ is the product of area ratio and the intensity contrast.\\

In this paper, we present a second hypothesis for the origin of the short flux-rise observed with HST, occurring if two planets transit simultaneously, and one of the planets occults the other: In a system of two transiting planets with different orbits, the known one has a short period (planet A) and the hypothetical other one has a much longer period (planet B). Assuming that the transit of the short-period planet, i. e. planet A, is being observed and that the centers of both transits coincide more or less in time, then planet B can be considered as moving slowly across the stellar disc. When planet A is eclipsing planet B; it is causing in this moment a flux increase in the combined transit light curve. Due to the short observation span, the ingress and egress of the much longer transit events of planet B might neither have been observed nor detected, see Figure \ref{model_TrES1_2_full} for an arbitrary example light curve. For a light curve of two transiting planets, the flux variation depends on the area ratio and on the impact parameter between both planets, whereas the duration depends on the area ratio and on the respective orbital velocities of both planets. \\

In Section 2 we describe the HST and IAC-80 observations and data reductions. We identified an increase in flux in one HST transit observation. Diagnostics to differentiate between the two origins of the flux rise are presented in Section 3, where we evaluate its color dependency from red to blue and then further investigate the flux variability of the other IAC-80 light curves. A discussion of the results will take place in Section 4.\\


\section{Observations and data reduction}
\label{reduction}

Table \ref{obs} gives an overview of the observations. The epochs refer to the planetary orbit; for the calculated mid-transit times we used the ephemeris from \citet{2004ApJ...613L.153A}, T$_c= 2,453,186.8060 + 3.030065 \times$ Epoch. Partial transits were observed in three 'visits' by the HST on Nov. 19 2004, Jan. 19 2005 and March 29 2005. These observations followed the protocol established by \citet{2001ApJ...552..699B}, where a complete transit light curve is obtained from phased overlay of data from several short 'visits' of the transiting system during slightly different phases. This phasing is necessary to overcome the interruptions due to Earth's occultation of the HST. For these HST observations the ACS/HRC instrument with the grism G800L was used. We collected spectra in the wavelength range between 5794 $\AA$ and 10251 $\AA$. The exposure time was 95 s.\\

Analysis of the HST images started with the pipeline-calibrated data, which subtracts zero point offset (based on over-scan mean), a two-dimensional bias and a dark (scaled by the exposure time) image. A wavelength dependent flat field, appropriate to position along the dispersion direction, has been derived by fitting a quadratic polynomial in wavelength to the eight broad-band flats over a range from 435 to 850 nm. A global sky level is computed by taking the median within a large region offset from the first order spectrum; this sky level was then subtracted from the images.  A total of 20 bad pixels (out of an extraction box totaling 10700 pixels) were flagged as bad and linearly interpolated over, before proceeding. Cosmic rays are eliminated by sigma-clipping over stacks of 108 images on a per pixel basis, taking into account fluctuations of the flux, which are correlated with minor x-y image motions and changes of the focus. The pointing offsets in x and y are evaluated by taking first moments of the (nearly point source) zeroth order image, and focus changes are tracked by fitting Gaussians in cross-dispersion along the spectrum and averaging them. For photometry summing over the whole first order spectrum sums are taken out to an isophote, at which the per image counts (in a mean image) have fallen to 220 electrons (compared to a peak of 108000). Sums in 500 $\AA$ bins use the same cross-dispersion limits, and contiguous blocks of columns most closely matching the desired ranges are formed. In addition a smaller bin about 62 $\AA$ wide is formed at H-alpha.\\

Summation over the full spectrum results in total counts of about 4.5$\times$10$^7$ electrons per exposure. Normalization is done by taking the mean over all points in orbits 2 and 5 and dividing by this. This sum spans wavelengths from 5392 to 10650 $\AA$. Similar normalizations are applied separately to the wavelength restricted bin time series. De-correlations are performed by evaluating a linear least-squares fit to points outside the transit as a function of x-y offsets, the cross-dispersion spectrum width and the time to remove any linear trends in each visit. The resulting function is evaluated at all points including inside the transit and then divided out.  Our final scatter in the time series of individual 500 $\AA$ bands is increased by about 9\% over the sum of Poisson and readout noise. Any residual band-to-band correlated noise must be relatively small given the close approach to being Poisson noise limited.\\

At eleven nights between 2004 and 2007, we also gathered rapid cadence photometry of TrES-1 transits with the IAC-80 telescope at the Observatorio del Teide, Spain. All observations were done in the Johnson R filter. In the first two years, we used a camera with a 1k $\times$ 1k CCD, whereas in 2006, it was replaced by a camera with a 2k $\times$ by 2k chip. The pixel scale of this new CCD is 0.305 \arcsec, resulting in a field-of-view of 10.25 x 10.25 \arcmin. Out of the eleven IAC-80 transit observations, only the seven that cover full transit events were useful for further investigations. Table \ref{obs} shows details of the observations used. \\        

\begin{table*}
\caption[]{Observing log.}
\label{obs}
\centering
\begin{tabular}{ccccccc}
\hline\hline
\noalign{\smallskip}
Day of         & Epoch & Calculated       & Telescope and  & Filter & Cadence & Standard deviation \\
observation    &       & mid-transit time & Instrument     &        & [s]	   & off-transit part\\
               &       & HJD-2,450,000    &                &        &         & [mmag] \\
\hline
 19. Nov. 04   & 47    & 3329.2191        & HST-ACS/HRC    & G800L  & 130   & 0.18\\
 19. Jan. 05   & 67    & 3389.8204        & HST-ACS/HRC    & G800L  & 130   & 0.18\\
 29. Mar. 05   & 90    & 3459.5119        & HST-ACS/HRC    & G800L  & 130   & 0.18\\
 10. Jul. 05   & 124   & 3562.5341        & IAC-80 old CCD & R      & 57    & 1.9\\
 16. Jul. 05   & 126   & 3568.5942        & IAC-80 old CCD & R      & 97    & 1.8\\
 08. Aug. 06   & 254   & 3956.4425        & IAC-80 new CCD & R      & 79    & 1.0\\
 11. Aug. 06   & 255   & 3959.4726        & IAC-80 new CCD & R      & 80    & 1.4\\
 16. Jun. 07   & 357   & 4268.5392        & IAC-80 new CCD & R      & 70    & 1.5\\
 19. Jun. 07   & 358   & 4271.5693        & IAC-80 new CCD & R      & 60    & 1.7\\
 22. Jun. 07   & 359   & 4274.5993        & IAC-80 new CCD & R	 & 70    & 1.7\\ 
\hline
\end{tabular}
\end{table*}

\begin{figure*}
\centering
\includegraphics[angle=0,scale=0.9]{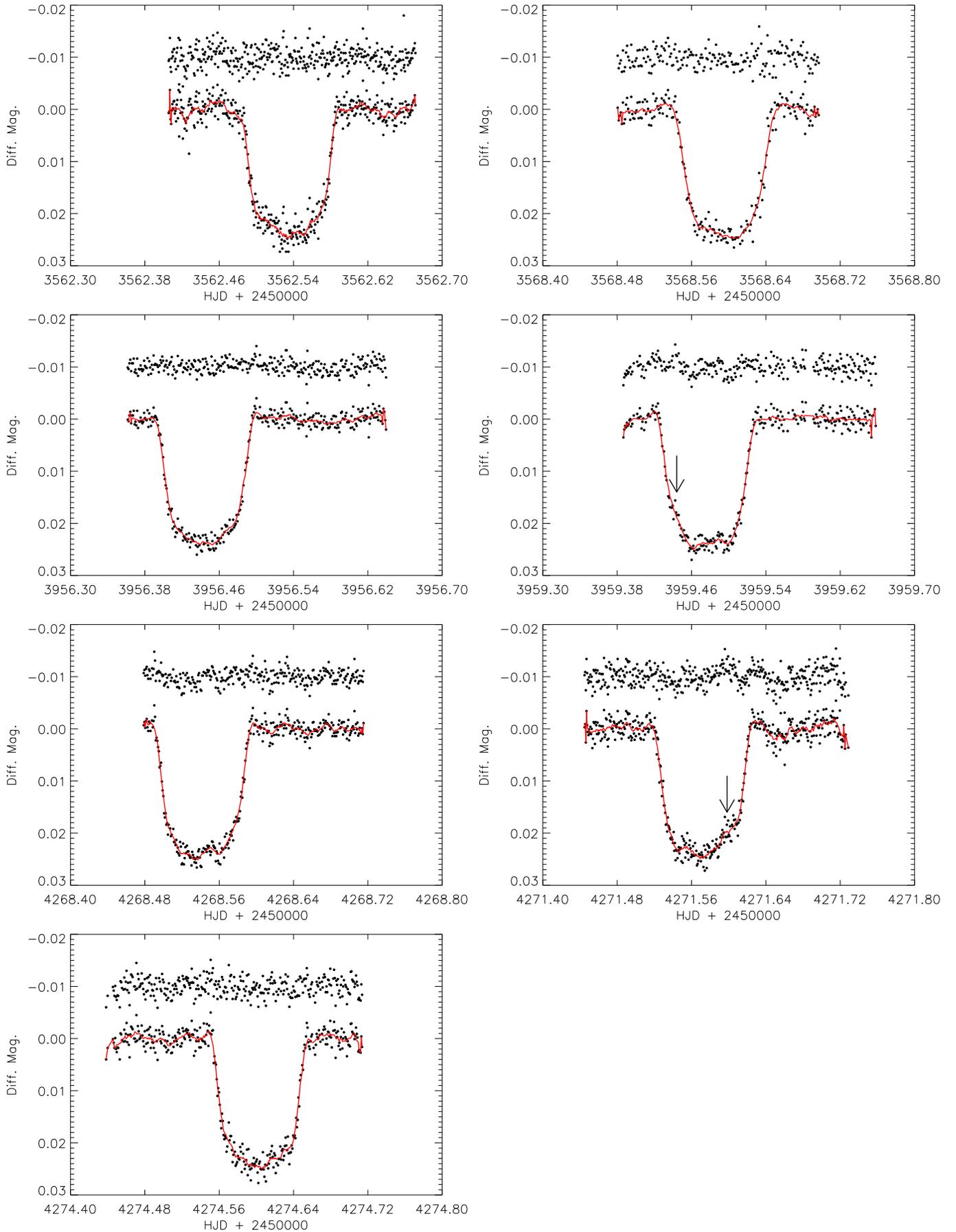}
\caption{Corresponding to Table \ref{obs}, IAC-80 observations of TrES-1 transits. The upper light curve is after subtracting a common transit model of TrES-1. The lower curve shows the observed light curve with the solid (red) line being a boxcar smoothing and the arrows indicating possible low significance detection of similar flux rises as seen in the HST data. \label{all_trans}}
\end{figure*}

The images of each night were calibrated using standard IRAF procedures. We carried out aperture photometry with {\tt VAPHOT} \citep{2001phot.work...85D} on the target and on five to eight comparison stars of similar brightness within the same CCD frame. In {\tt VAPHOT}, differential photometry is obtained by dividing the target flux by the reference flux, which was constructed from a weighted ensemble of comparison stars. The flux was then converted into magnitudes and corrected for differential extinction by subtracting a parabolic fit to the off-transit parts of each individual light curve. We estimated the photometric error using the standard deviation outside the transit, see Table \ref{obs}. Figure \ref{all_trans} shows the transit light curves of TrES-1 observed with the IAC-80.\\

\begin{figure}
\centering
\includegraphics[angle=90,scale=0.34]{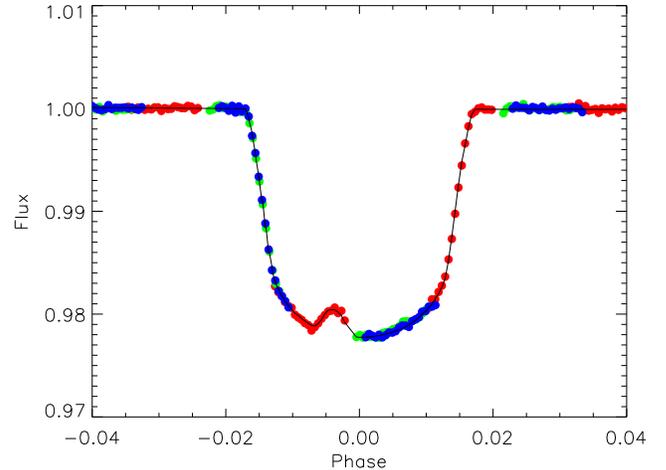}
\caption{Phased light curve of all three HST visits using first order spectrum, each color represents a different day, i. e. red corresponds to November 19, 2004 , green to January 19, 2005 and blue to March 29, 2005. A flux increase is seen in the transit light curve on November 19, 2004. The black solid line is an example of a model for a simultaneous transit of TrES-1 and an arbitrary second transiting planet using UTM (see text Section \ref{DiscCon}). In this example, the second planet has an arbitrary period of 300 days and radius of 4.25 R$_{\oplus}$. A nearly grazing impact between TrES-1 and the second planet was simulated. \label{HST_visits}}
\end{figure}

By visually inspecting the phased HST light curve (see Figure \ref{HST_visits}), we can identify an increase in flux during the transit on the night of November 19, 2004. The feature has a height of 2.7 mmags, whereas the standard deviation out of transit is 0.18 mmags. Hence, this feature cannot be attributed to noise. The feature also is highly unlikely to be caused by instrumental effects. No correlated variations in x-y pointing or focus are present at the time of the flux rise, i.e. nothing suggestive of an instrumental effect. During the several HST orbits, outside of the transit no features remotely similar to this have ever been seen. The short flux increase during the TrES-1 transit is $\sim$ 10 min. long. This correspond to 1/15 of the whole transit of TrES-1. This kind of flux increase can occur when TrES-1, during the transit, passes over a darker area on the stellar surface. 


\section{Analysis of the observed flux rise}
\label{Analysis}

The two hypotheses presented previously, e. g. that the darker patch is caused by a) a cooler starspot or b) a second transiting planet would differ by these properties:

\begin{itemize}
\item A starspot should show a flux-rise with a wavelength dependency, with a higher flux rise in the blue than in the red wavelength range. A flux rise from a mutual planet-planet occultation should not have any significant wavelength dependency.\\
\item A starspot is a temporary phenomena on time-scales of weeks to months; spots appear and disappear with different sizes at different positions. The planet-planet occultation would be a very rare event, except if the two planets are in a resonant orbit.\\
\end{itemize}

\begin{figure}
\centering
\includegraphics[angle=0,scale=0.4]{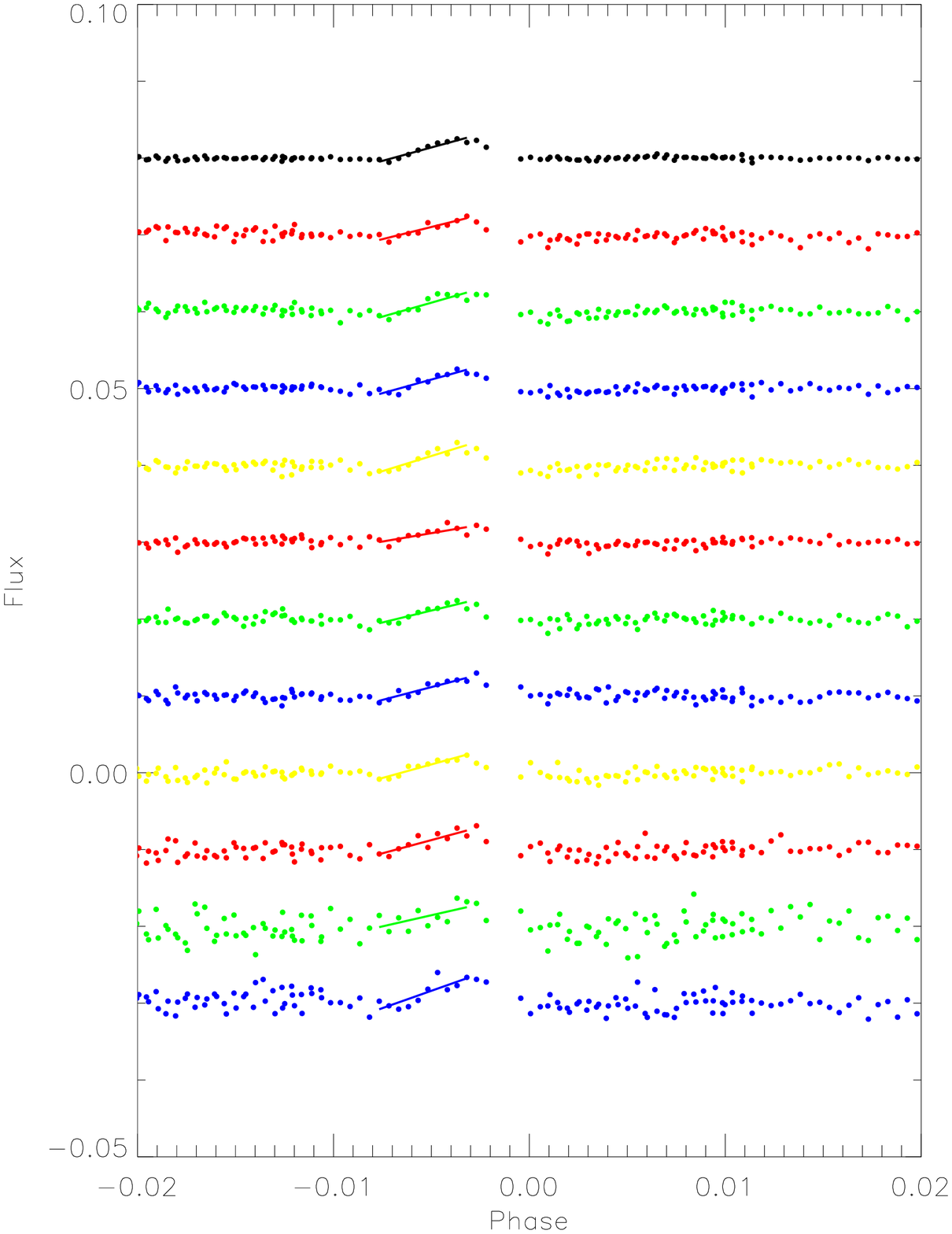}
\caption{Phased light curve of all HST observations with the best fit transit model subtracted. From the bottom to the top are the light curves in the bins with central wavelength of H-$\alpha$, 10251, 9753, 9247, 8745, 8245, 7750, 7260, 6762, 6250, 5790 $\AA$. At the top is the sum over all wavelengths. The short inclined solid lines indicate the slope that was fitted to the flux rise. \label{HST_modelsub}}
\end{figure}

We checked if there is any wavelength dependency of the flux rise feature during the transit of TrES-1. To this end, we created an HST transit light curve model by using the phased light curve of all three first order spectra HST observations and assigned a zero statistical weight to the data points of the feature, i. e. the flux rise was not considered for the following modelling. For transit modelling we used the FORTRAN routines from \citet{2006A&A...450.1231G} and the simplex-downhill fitting algorithm \citep{1992nrfa.book.....P}, minimizing $\chi^2$. The free parameters were the ratio between the stellar and planetary radii, $k=R_{p}/R_{*}$, the sum of the projected radii, $rr= (R_{*}/a)+(R_{p}/a)$, and the limb darkening coefficient $u_1$ and $u_2$. Next, we used the color bins of the HST grism spectra and fitted the limb darkening coefficient in each wavelength bin, keeping $rr$ and $k$ constant, as obtained by the best fit of the first order spectra. The residuals are plotted in Figure \ref{HST_modelsub}.  \\

\begin{figure}
\centering
\includegraphics[angle=90,scale=0.34]{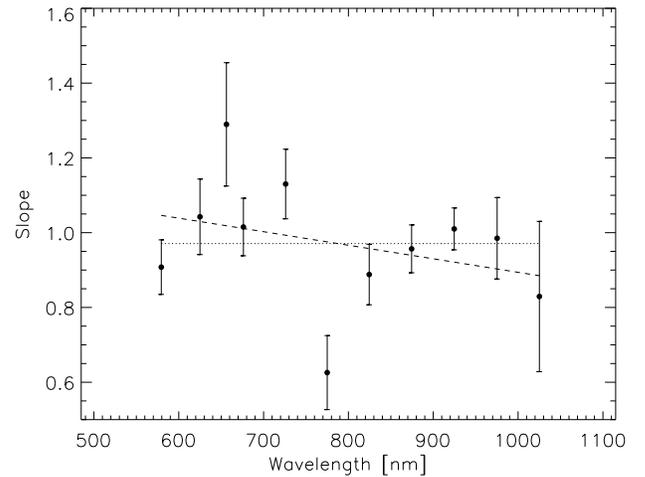}
\caption{Values of normalized fitted slopes for the flux rise in each wavelength bin and corresponding error bars. The dashed line is a two parameter linear fit and the dotted line a one parameter linear fit. The two parameter linear fit indicates a wavelength dependency and the one parameter fit a independence.\label{HST_slope}}
\end{figure}

We used the model-subtracted light curves and estimated the slope of the flux increase in each wavelength bin following the procedure by \citet{2007A&A...476.1347P}, see Figure \ref{HST_modelsub}. The estimated slopes were then normalized with the slope estimated from the summed light curve. Figure \ref{HST_slope} shows the estimated normalized slopes for each wavelength bin and their corresponding error bars. The slope in the H-$\alpha$ bin is a little higher; if real, this would indicate that TrES-1 occults a H-$\alpha$ rich region during transit. However, the light curve in H-$\alpha$ is also noisier and its error bars are higher, indicating that its higher slope might not be real. The slope of the bin with the central wavelength of 7750 $\AA$ is lower, but we attribute this to a single outlier. Also the bin centered at 10251 $\AA$ has high error bars due to a noisier light curve.\\

We used the above estimated slope values and fitted twice a linear polynomial of the form $y=A+B \lambda$ to them, where $\lambda$ is the central wavelength of the bin and y the normalized slope. In the first fit, we fitted for both parameters, $A$ and $B$, and in the second one, we kept $B$ at zero and fitted for $A$ only. The first case indicates a wavelength dependency, the second case indicates a wavelength independence of the flux rise. For the first case we obtained a reduced $\chi^2$ of 2.8 and for the latter case a reduced $\chi^2$ of 2.3. After applying the F-test, we found that there is no significant difference between both hypothesis, wavelength dependency and independence, within a 5\% confidence interval. \\

\begin{figure*}
\centering
\includegraphics[angle=0,scale=0.9]{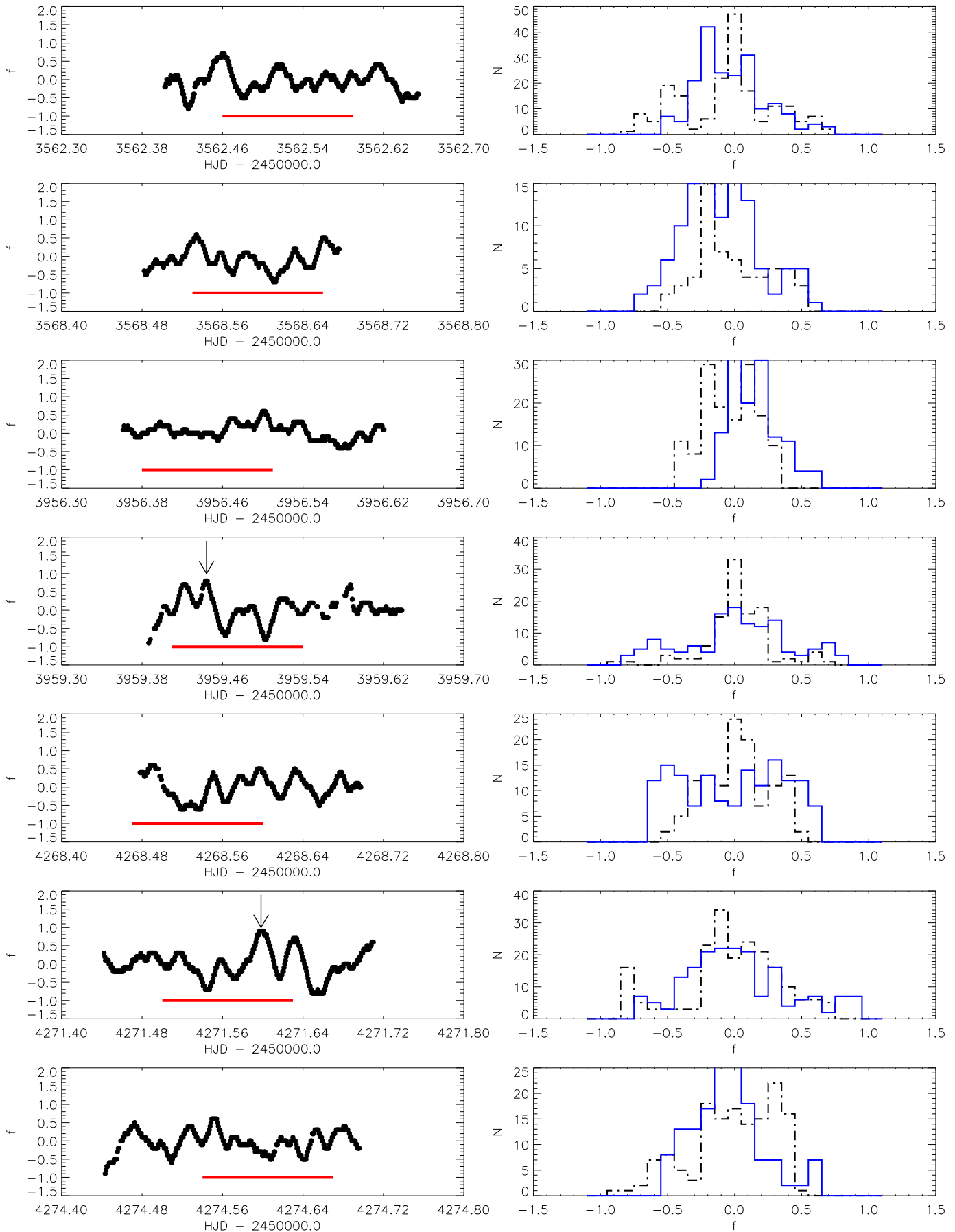}
\caption{Left side: Best fit $f$ for each transit, where $f$ is a multiplication factor of a model template, see text in Section \ref{Analysis}. The solid (red) line below indicates the duration of the transit of TrES-1 and the arrows indicate a possible low-significance detection of a similar feature in the IAC-80 data. Right side: Histograms of $f$. The dotted lines give histograms for the out-of-transit part, the solid (blue) lines represent histograms of the transit part. \label{spot_simu}}
\end{figure*}

Furthermore, we used our IAC-80 data to check if these observations contain the same or similar features of flux-rises during transit. However, the IAC-80 observations are noisier and hence, similar features cannot be seen as easily as in the case of the HST light curve. Therefore, we created a best fit model for the transits from the IAC-80 observations, using the same approach as before, and subtracted the respective model. Then we used the flux rise event of the HST light curve as a template and applied a match filter detection against this event to the model subtracted IAC-80 observations, as follows: We shifted the template in time along the model subtracted IAC-80 light curves, fitting in each shift for $M_{fit}=M_{HST}*f$, where $M_{HST}$ is the magnitude of our template and $f$ is a multiplication factor, which is also the fitting parameter. The cadence of the HST observations is lower than of the IAC-80 and, hence, we interpolated our template, using a spline interpolation. The best fit for $f$ over time can be seen on the left side in Figure \ref{spot_simu} and a histogram of $f$ is shown on the right side. In the histograms, the value of $f$ should concentrate around 0 if there are no features, while a value close to 1 may mean a detection of a similar feature than observed with HST. Taking the off-transit standard deviation of $f$ with $\sim 0.27$ into account, we consider as candidates for the detection of such a feature values of $f \ge 0.8$. Inspecting the histograms in Figure \ref{spot_simu}, we identified two low significant events on HJD-245000 = 3959.41 and 4271.60. These events are also indicated with arrows in Figures \ref{all_trans} and \ref{spot_simu}.\\

%

\section{Discussion and Conclusions}
\label{DiscCon}

We observed transits of TrES-1 in 2004 and 2005 with the HST and identified an intriguing flux rise during one transit. We further followed-up transits of TrES-1 during the years 2004 to 2007 from the ground, finding two further possible flux rises, and used the color information in the HST data to assess the origin of this feature. We can identify two possible causes, one is the possibility of a circular starspot occulted by TrES-1 during transit and the other is the detection of a second smaller transiting planet in an orbit larger than TrES-1. \\

A good fit for the feature's shape can be obtained by using a circular dark patch on the stellar surface and neglecting deformation of a circular disk due to projection on a sphere. Assuming further that the dark patch and TrES-1 have a central impact on each other and using the orbital velocity for TrES-1 of $v=140$ km/s, then, in a time interval of 10 min., TrES-1 crosses 84000 km, which gives us a lower limit for the diameter of the occulted patch. The real diameter depends on the stellar rotation (for the starspot hypothesis) or on the orbital velocities of the transiting planets (for the planet-planet hypothesis). \\

The host star of TrES-1 is an active star \citep{2004ApJ...616L.167S} which is favouring the spot scenario. In that case, we calculated first the spot radius. Since the parent star of TrES-1 is slowly rotating with a rotational velocity of $I_S=1.08$ km/s \citep{2005ApJ...621.1072L}, corresponding to a rotation period of 38.4 days, the stellar rotation can be neglected and under the assumption of a central impact between TrES-1 and spot, the spot radius is simply half the diameter as obtained previously, i. e.  42000 km. An effective temperature of the stellar surface of 5250 K was obtained by \citet{2004ApJ...616L.167S} and combined with the equation 1 given by \citet{2003ApJ...585L.147S}, we estimated the spot temperature to be approximately 4690 K for a starspot with the radius of 42000 km. These values are perfectly reasonable for a large sunspot.\\

To generate a realistic model that shows that the observed HST transit light curve might as well arise from two occulting planets, we used ``Universal Transit Modeller'' (UTM, \citeauthor{2008arXiv0807.3915D} 2008). UTM is an IDL-routine permitting simulations of transit light curves of multiple bodies, including rings and moons, and the fitting of observed light curves. We fitted first the summed HST light curve without considering the flux rise part. As fit parameters we used the radii of the host star and of the planet TrES-1, respectively; the limb darkening coefficient, assuming a quadratic limb darkening law, the inclination and the distance. We then used the best fit parameters and introduced a second transiting planet, but in an expanded orbit. However, we lack the determination of an unique period for the putative second transiting planet due to the singularity of the observed event and we further have no significant evidence in the IAC-80 data to derive a period. \\

\begin{figure}
\centering
\includegraphics[angle=90,scale=0.34]{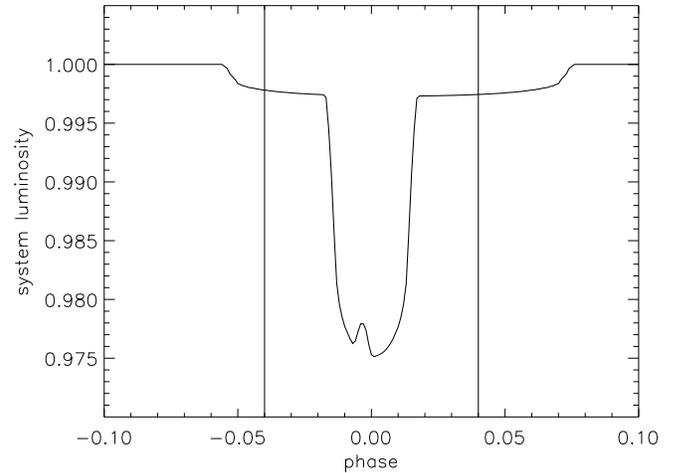}
\caption{Example of the same planet-planet transit model as shown in Figure \ref{HST_visits}, for TrES-1 and an arbitrary additional transiting planet, but showing the model over longer time span, so that the ingress and egress of a possible second planet are visible. The vertical solid lines indicate the usual observation span of 6 hours centered at the mid-transit of TrES-1. We note that during this 6 hour time span the apparent transit depth of TrES-1 is identical with or without a second planet.  \label{model_TrES1_2_full}}
\end{figure}

Hence, we can only give as an example a model of two transiting planets, see Figures \ref{HST_visits} and \ref{model_TrES1_2_full}. Figure \ref{model_TrES1_2_full} shows a complete example model, including ingress and egress of an arbitrary putative second planet. The introduced second planet has an arbitrary period of 300 days and radius of 4.25 R$_{\oplus}$. Its transit duration is 10 hours and the depth is 2.7 mmags. It is very unlikely that a complete transit of this second planet would have been observed, because the transit duration is too long for nightly observations and it is even more difficult to keep a sufficient high photometric precision from ground over such a long time span due to changing atmosphere, extinction and airmass. All IAC-80 observations were done in time spans shorter than 10 hours and even under ideal conditions, i. e. at low airmass, the differential photometry applied to our transit observations did not correct completely for extinction. Therefore, as noted in Section \ref{reduction}, we corrected the extinction by subtracting a second order polynomial fit from the observed light curves which removed all possible hints to a second planet's transit from the light curve's off-transit part.\\

Whereas a mutual occultation would be an extremely rare event, limits for the presence of transits from a second planet can be derived from size-estimates of this planet. Its minimum size can be obtained from the amplitude of the mutual transit event and assuming that TrES-1 and the other planet have a central impact on each other. In this case, the flux increase is directly related to the area of overlap of the projection of the two planets into the star, and is independent of the period. For TrES-1 we have a transit depth of 23 mmags and the flux rise height is 2.7 mmags, and hence we obtain an area of overlap of $1/8.5$ of the cross section of TrES-1. Knowing the radius of TrES-1, 1.081 R$_J$ \citep{2007ApJ...657.1098W}, we obtained a minimum size of a second planet of 0.37 R$_J$. However, the shape of the flux rise indicates a more grazing impact between TrES-1 and the hypothetical second planet, favouring a bigger radius.\\ 

The original STARE and PSST data of the field where TrES-1 has been discovered cover a time span of 96 days \citep{2004ApJ...613L.153A}. This is a time span long enough to possibly detect a putative second planet on a longer orbit, but no transit has been detected in this data. Therefore, based on an assumption of a standard deviation of 10 mmags in the TrES-1 discovery data, we can set an upper limit for the size of a possible second planet. In the limiting case, a transit with a depth of 10 mmags would not have been detected. This 10 mmags corresponds to a transit depth of a planet with a radius of 0.72 R$_J$, which constitutes an upper detection limit for the radius of a possible second planet. \\
\\
For a putative long-period outer planet and due to the limited and gaped observations of the TrES-1 field the detection depends mostly on the probability of having observed a transit. Therefore, we further constrain the period in which this upper radius limit is valid. We applied a statistical analysis to a time vector which contains the time of observation. We established 2000 trial periods between a period of 4 days and 100 days on the STARE/PSST discovery data. Within each trial period we set randomly 300 epochs and tested how often we could recover a transit signal. We set a detection limit for a possible planet at periods where the recovery rate drops below 95 \%. For the original STARE and PSST data this occurs at a period of 10.5 days. Hence, for periods of less than 10.5 days, we can set an upper limit for the radius of 0.72 R$_J$ for a possible second planet. However, larger planets in longer periods are possible.  \\
\\
Based on the conventional interpretation of flux-rise features during transits, we may conclude that a starspot has been detected in the HST data and possibly as well in the ground-based IAC-80 data. We also presented an alternative hypothesis of a mutual planet-planet occultation, which would be a  rare but not impossible event. The verification or exclusion of this hypothesis is difficult unless a wavelength-dependency of the flux-rise features can be demonstrated, or until the presence of the secondary planet can either be verified or excluded, be it from observations of its own transits or from radial velocity measurements.

\begin{acknowledgements}
This article publishes observations made with the IAC-80 telescope operated by the Instituto de Astrof\'isica de Canarias in the Observatorio del Teide. MR likes to thank the European Association for Research in Astronomy (EARA) for their support with an EARA - Marie Curie Early Stage Training fellowship. Some of the analysis described here is also based on data from the STARE and PSST telescopes. We also thank the anonymous referee for valuable comments.
\end{acknowledgements}

\bibliographystyle{aa} 
\bibliography{bib_mar}

\end{document}